\documentclass[11pt,onecolumn,a4paper]{quantumarticle}

\newif\ifARXIV
\newif\ifUSEBBL

\ARXIVtrue

\USEBBLtrue

\pdfoutput=1
\usepackage[font={small}]{caption}
\usepackage{subfigure}
\usepackage{graphicx}
\usepackage[numbers,sort&compress]{natbib}
\usepackage{float}
\usepackage[normalem]{ulem}
\usepackage{mathrsfs} 
\usepackage{multicol}
\usepackage{cancel}
\usepackage{mathtools}
\usepackage{hyperref}
\usepackage{xcolor}
\usepackage{simplewick}
\usepackage{amsmath}
\usepackage{amsfonts}
\usepackage{amssymb}
\usepackage{graphicx}
\usepackage{simplewick}
\usepackage{hyperref}
\usepackage{multirow}
\usepackage{enumitem}
\usepackage{physics}
\usepackage{booktabs}
\usepackage{xcolor}
\usepackage{dsfont}
\usepackage{subfigure}
\usepackage{amsmath,amsthm,appendix,bm,graphicx,hyperref,mathrsfs,setspace,slashed,xcolor}
\usepackage{quantikz,stmaryrd}
\def\Z2{ {\mathbb Z}_{2} }

\newcommand{\BRA}[1] {\langle  #1 |}

\newcommand{\KET}[1] {| #1 \rangle}

\def\ALLNQI{PhysRevLett.81.3992,150706334,BechmannPLA98,9802051,9803019,9811036,0309189,0502072,BrunPRL09,BennettPRL09,13033537,13030371,13107301,200907800,220613362,XuPRR22,211105977,240416288,240310102,250306423}

\begin{document}


\title[]{Nonlinear Hamiltonians and Boolean satisfiability}

\author{Michael R. Geller}
\email{mgeller@uga.edu}
\affiliation{Department of Physics and Astronomy, University of Georgia, Athens, Georgia 30602, USA}
\affiliation{Center for Simulational Physics, University of Georgia, Athens, Georgia 30602, USA}

\author{Victoria S. Ordonez}
\email{vordonez@uga.edu}
\affiliation{School of Computing, University of Georgia, Athens, Georgia 30602, USA}
\affiliation{Center for Simulational Physics, University of Georgia, Athens, Georgia 30602, USA}

\author{Yohannes Abate}
\email{yohannes.abate@uga.edu}
\affiliation{Department of Physics and Astronomy, University of Georgia, Athens, Georgia 30602, USA}

\date{May 14, 2026}

\begin{abstract}
We consider an extended model of quantum computation where a scalable fault-tolerant quantum computer is coupled to and entangled with one or more ancilla qubits that evolve according to a nonlinear Schr\"odinger equation. Following the approach of  Abrams and Lloyd, an efficient quantum circuit evaluating an $n$-bit  Boolean function in conjunctive normal form is used to prepare an ancilla encoding its number $s$ of satisfying assignments ($0 \! \le \! s \! \le \! 2^n$). This is followed by a nonlinear quantum state discrimination gate on the ancilla qubit that is  used to learn properties of $s$. Here we consider three types of state discriminators generated by different nonlinear Hamiltonians. First, given a restricted Boolean satisfiability problem with the promise of at most one satisfying assignment  ($ 0 \!  \le \!  s \!  \le  \! 1$), we show that a qubit with $\langle \sigma^z \rangle \sigma^z$ nonlinearity  can be used to efficiently determine whether $s \! = \! 0$ or $s \! = \! 1$, solving the UNIQUE SAT problem. Here  $\langle A \rangle\!  := \!  \BRA{\psi} A \KET{\psi} $ denotes expectation in the current state. UNIQUE SAT is {\sf NP}-hard under a randomized polynomial-time reduction (of course any discussion of complexity assumes a scalable, fault-tolerant implementation). Second, for unrestricted satisfiability problems with $ 0 \! \le \!  s \!  \le \! 2^n$, a Hamiltonian with  $ \langle \sigma^x \rangle \sigma^y  \! - \! \langle \sigma^y \rangle \sigma^x$ nonlinearity can be used to efficiently determine whether $s=0$ or $s>0$, thereby solving 3SAT, which is {\sf NP}-complete. Finally, we show that $ \langle \sigma^y \rangle \langle \sigma^z \rangle  \sigma^x \! - \! \langle \sigma^x \rangle \langle \sigma^z \rangle  \sigma^y $ nonlinearity can be used to efficiently measure $s$ and solve \#SAT, which is \#{\sf P}-complete. These models allow us to explore the interplay between distinct forms of nonlinearity and their resulting computational power, at least in the noise-free limit. The nonlinear models are of mean field type and might be simulated with ultracold atoms.
\end{abstract}

\maketitle
\setcounter{tocdepth}{2}
\clearpage
\tableofcontents


\section{Introduction}

The linearity of the Schr\"odinger equation 
imposes significant restrictions on the power of quantum computers by making any 
deterministic nonlinear operation imperfect (impossible to implement with perfect fidelity even in the noise-free limit).
The most prominent example is the impossibility of cloning a quantum state \cite{WoottersZurekNat1982}.
Another is the impossibility of increasing the orthogonality or distinguishability of a pair of 
nonorthogonal states $\rho_{\rm a}$ and $\rho_{\rm b}$.  This is captured by the trace distance monotonicity condition \cite{RuskaiRMP1994,WildeQuantumInformation}
\begin{equation}
\|  \rho_{\rm a}^{\prime}   - \rho_{\rm b}^{\prime}   \|_1\ \le \| \rho_{\rm a} - \rho_{\rm b} \|_1 ,
\label{trace monotonicity}
\end{equation}
satisfied by any linear positive trace-preserving
(PTP) map $\rho \mapsto  \rho^{\prime}$, with the equality applying  to unitary maps. 
Here $ \| X \|_1 := {\rm tr}(|X|)$ is the trace norm,
where $ |X| =\sqrt{ X^\dagger X}.$
Applied to a qubit with Bloch vector ${\vec r} = (x,y,z) = {\rm tr}(\rho {\vec \sigma} )$,
the condition (\ref{trace monotonicity}) 
reduces to
\begin{equation}
| {\vec r}_{\rm a}^{\, \prime} - {\vec r}_{\rm b}^{\, \prime} |
\le | {\vec r}_{\rm a} - {\vec r}_{\rm b}  | .
\label{trace monotonicity qubit}
\end{equation}
This means that, under the same linear map, qubit states can move closer together but never further apart, limiting the ability to distinguish between them after subsequent measurement.
As a consequence, it is impossible to implement perfect quantum state discrimination, where a pair of potential inputs $( \KET{a}, \KET{b}) $
map to perfectly distinguishable classical states  $( \KET{0}, \KET{1}) $ \cite{08101970,12042313,BaeJPA15,210812299}.
  
This makes it interesting to consider extended models of quantum computation where these restrictions are bypassed \cite{\ALLNQI}.
 Aaronson has argued that a noise-free quantum computer with certain nonlinear AND and OR gates could solve any {\sf PSPACE} problem in polynomial time
 \cite{0502072}, where {\sf PSPACE} 
 is the set of decision problems that can be solved using a polynomial amount of memory with a deterministic Turing machine.
 Here  we explore this question from the other direction, investigating the computational power as 
 increasingly complex single-qubit nonlinearity is included. We do this by designing three quantum state discrimination gates, each generated by a different nonlinear Hamiltonian, and applying them to the problem of Boolean satisfiability \cite{AroraBarak2009}. This is a useful setting for studying computational complexity because it sits at the center of many theoretical and practical problems in computer science.
 
 In the next Section we develop a simple method for constructing nonlinear Hamiltonians. 
 In  Section~\ref{amplitude encoding section} we review the approach of Abrams and Lloyd \cite{PhysRevLett.81.3992}  to encode  the number of solutions $s$ to a Boolean satisfiability problem into the probability amplitudes of an ancilla qubit.
 In the remainder of this paper we design nonlinear Hamiltonians and quantum state discrimination gates to measure properties of $s$, and ultimately its precise value. In Section~\ref{torsion model section} we construct a nonlinear Hamiltonian 
 (the $z$-axis torsion model 
\cite{MielnikJMP80,150706334,241022032}) to efficiently determine whether $s \! = \! 0$ or $s \! = \! 1$, solving the UNIQUE SAT problem. 
In Section~\ref{morse–smale section} we design a Hamiltonian to efficiently determine
whether $s=0$ or $s>0$, thereby solving 3SAT.
In Section~\ref{pitchfork section} we design a Hamiltonian to efficiently measure the precise value of $s$, solving \#SAT. 

\subsection{Nonlinear velocity fields on the sphere}

We approach the design of nonlinear Hamiltonians and gates as an inverse problem, starting with some desired phase portrait for the qubit---including locations and types of fixed points---and working backwards to find the associated nonlinear Hamiltonian.  The first step is to  construct a velocity field $ {\vec v} = {\vec v}({\vec r}) $ on the Bloch sphere that gives the desired phase portrait,  while also conserving $ |{\vec r}| $ (we consider only pure states in this paper).
Velocity fields that are linear in 
${\vec r}$ result in linear qubit models, whereas nonlinear velocity fields result in nonlinear, state-dependent Hamiltonians. 

To conserve $ |{\vec r}|$,  we require 
 $ {\vec v}  \cdot {\vec r} =0$.
Any tangential velocity field on the unit sphere can be written in terms of a second field 
$ {\vec u} = {\vec u}({\vec r}),$ 
orthogonal to ${\vec v}$, as 
\begin{equation}
{\vec v}  = {\vec u}  \times {\vec r} , \ \ 
{\vec v}  \cdot  {\vec u}  = 0, \ \ 
{\vec v}  \cdot  {\vec r}  = 0, \ \ 
 |{\vec r}| = 1.
 \label{tangent field decomposition}
\end{equation}
The ${\vec u}$ field determines the linear or nonlinear Hamiltonian 
\begin{equation}
H =  \frac{  {\vec u} \big(   \langle  {\vec \sigma} \rangle  \big) 
 \cdot  {\vec \sigma} }{2}  
 \label{general nonlinear hamiltonian}
 \end{equation}
 that has $ {\vec v} $ for its  Bloch vector 
 equation of motion:
$ \! \frac{ d {\vec r}}{dt} = {\vec v} $.
To see this, use
$ \rho = \frac{1}{2}(I + {\vec r} \cdot {\vec \sigma}) $
and
\begin{equation}
\frac{ d\rho}{dt} = -i [H, \rho] = -\frac{i}{2}  
u^\mu (\langle  {\vec \sigma} \rangle) \,  \big[
 \sigma^\mu , \rho \big]  = \frac{1}{2}  
 \big( {\vec u}({\vec r}) \times {\vec r} \, \big) \! \cdot {\vec \sigma} ,
 \end{equation}
where $I$ is the two-dimensional identity and
 the $\sigma^\mu \, (\mu \! \in \! \{ x,y,z \})$ are Pauli matrices. Then  $ \frac{d {\vec r}}{dt} = {\rm tr}( \frac{ d\rho}{dt} {\vec \sigma}) = {\vec u} \times {\vec r}$ as required. 

The forward direction of the design problem,
finding ${\vec v}$ given ${\vec u}$ and  $H$, is 
immediately solved 
using (\ref{tangent field decomposition})
and (\ref{general nonlinear hamiltonian}).
However our main interest is the inverse problem
of determining  ${\vec u}$  and $H$ from 
 ${\vec v}$. In the case we take
\begin{equation}
 {\vec u} :=   {\vec r} \times {\vec v} , \ \ 
{\vec u}  \cdot  {\vec r}  = 0, \ \ 
{\vec u}  \cdot  {\vec v}  = 0, \ \ 
 |{\vec r}| = 1.
 \label{u field}
\end{equation}
The only difference from the forward direction is that now, by construction,  ${\vec u} $
is automatically tangential: ${\vec u}  \cdot  {\vec r}  = 0$. As we can see from the equation of motion
${\vec v}  = {\vec u}  \times {\vec r} $, the presence of a radial component of ${\vec u}$ ,
such as $\Lambda({\vec r}) \, {\vec e}_r$, does not
affect ${\vec v}$. 

In the gate design problem for pure states, any radial component of ${\vec u}$ is a gauge degree of freedom that changes the form of the Hamiltonian, but not the equation of motion. 
To see this, suppose that ${\vec u}$ is purely radial: ${\vec u}({\vec r}) = \Lambda({\vec r}) \, {\vec e}_r.$
In this case we expect there to be no dynamics of the qubit
state or Bloch vector.  Using $| {\vec r} | =1$ we obtain
\begin{equation}
 \frac{ d\rho}{dt} = -i [H, \rho] = 
-i \left[ \Lambda({\vec r}) \,  \frac{{\vec r} \cdot {\vec \sigma} }{2}   , \rho \right] 
= -i  \Lambda({\vec r}) \left[  \rho - \frac{I}{2}  , \rho
\right]  = 0.
 \end{equation}
Although it is always possible to use a purely tangential ${\vec u}$ field (for example, by subtracting a radial component), this does not always lead to the simplest Hamiltonian. 

In the linear case, ${\vec u} $ is a
uniform ($ {\vec r} $ independent) field, 
and  the dynamics
(\ref{tangent field decomposition})
reduces to ordinary precession of the Bloch vector about  the fixed vector ${\vec u} $. In the nonlinear generalization, each Pauli matrix 
$\sigma^\mu $
is multiplied by an independent function 
of the  Bloch vector
$ {\vec r} \! = \! \langle  {\vec \sigma} \rangle $.
Similar approaches have been used 
in a variety of contexts \cite{NikuniWilliamsJLTP2003,FarolfiZenesiniPRA2021,FarolfiZenesiniNatPhys2021,240219429}.

An additional simplification occurs when
the surface flow described by the velocity field 
is time-independent and incompressible, meaning that area elements move under ${\vec v}$ without expanding or contracting. Incompressibility in the present context requires that there are no local sources or sinks. Let  
\begin{equation}
{\rm div}_{S} \, {\vec v} 
= {\textstyle{\frac{1}{\sin\theta}}}
\left[  \partial_\theta 
  (v_\theta  \sin\theta  )  +
\partial_\phi  v_\phi  \right]
  \label{intrinsic div}
 \end{equation}
be the divergence of $ {\vec v}$ 
intrinsic to the surface of the Bloch sphere, and let
\begin{equation}
{\rm curl}_{S} \, {\vec u} 
= {\vec e}_r \cdot ( \nabla \times {\vec u} ) = 
{\textstyle{\frac{1}{\sin\theta}}}
\left[   \partial_\theta  ( u_\phi \sin\theta ) -   \partial_\phi u_\theta \right] 
\label{intrinsic curl}
 \end{equation}
be the intrinsic curl of $ {\vec u}$. 
Using
 ${\vec u} = {\vec e}_r \times \left( v_\theta {\vec e}_\theta + v_\phi {\vec e}_\phi \right) = v_\theta {\vec e}_\phi - v_\phi {\vec e}_\theta $
 leads to 
 \begin{equation}
  \operatorname{div}_S {\vec v} =
 \operatorname{curl}_S {\vec u}.
  \label{curl div identity}
 \end{equation}
 Incompressibility, namely $\operatorname{div}_S {\vec v} = 0$, therefore implies
  \begin{equation}
 {\rm curl}_S \, {\vec u} = 0.
 \label{curl free u}
 \end{equation}
Condition (\ref{curl free u})  always applies in the linear unitary case
 because ${\vec u}$ is uniform in that case.
It implies that there exists a scalar potential $E({\vec r})$, defined on the sphere, such that
${\vec u}  \propto  \nabla E$:
 \begin{equation}
{\vec u}  = 2 \, \nabla E \ \ 
{\rm and} \ \ 
\frac{dE}{dt} = \nabla E \cdot \frac{d {\vec r} }{dt}
= \frac{1}{2} \, {\vec u}  \cdot  {\vec v} = 0.
 \label{energy conservation}
 \end{equation}
The second condition in 
(\ref{energy conservation}) shows that the energy $E$ is a constant of the motion. The conditions ${\vec u}  \propto  \nabla E$ and $dE/dt=0$ uniquely specify $E$ up to a multiplicative scale factor (and additive constant). The scale factor can be chosen so that, in the linear case, the conserved energy is the expectation of $H$:
 \begin{equation}
  E({\vec r}) = \langle H \rangle = \frac{\langle {\vec u} \cdot  \vec \sigma \rangle }{2} 
  = \frac{{\vec u} \cdot {\vec r}}{2}  . \ \ 
  ({\rm linear \ and \ unitary})
 \label{energy conservation linear case}
 \end{equation}
This correspondence leads to the 
 factor of two in 
 (\ref{energy conservation}).
 
\subsection{Amplitude encoding}
\label{amplitude encoding section}

Next we explain the approach of Abrams and Lloyd \cite{PhysRevLett.81.3992} for encoding  the number of solutions $s$ of a Boolean satisfiability problem into the probability amplitudes of a single qubit.  Each problem instance is specified by a Boolean function $f : \{0,1\}^n \rightarrow \{0,1\} $ on $n$ bits, in conjunctive normal form. 
In particular, $f(x)$ is the logical AND of ${\rm poly}(n) \!$ clauses, with each clause the OR of three variables or their negations (this is sufficient for {\sf NP}-completeness \cite{AroraBarak2009}).
It is also assumed that the classical evaluation of $f(x)$ is efficient.
A unitary $U$ (also efficient) 
implements the function
$f(x)$ according to 
 $U \KET{x} \KET{y} = \KET{x} \KET{y \oplus \! f(x)}$. Here $\oplus$ denotes addition mod 2. 
 
\begin{figure}
\begin{center}
\includegraphics[width=11.0cm]{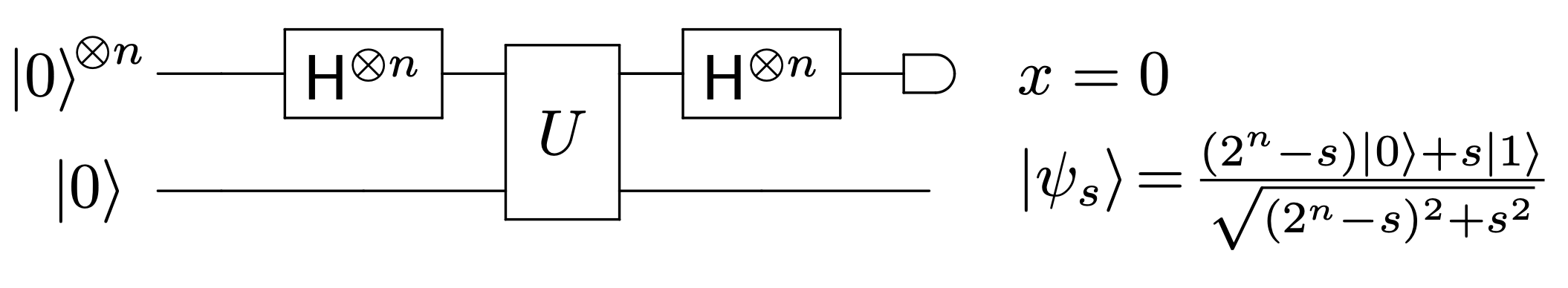} 
\caption{Efficient preparation of an ancilla qubit encoding the number of solutions $s$ to a Boolean satisfiability problem. The first (upper) register contains $n$ qubits storing the bitstrings $x \in \{ 0,1\}^n$, while the second register consists of  a single ancilla qubit. Here ${\sf H}$ is a Hadamard gate and $U$ implements the Boolean function
$f(x)$. After completion of the circuit, the first register is measured  and postselected to $\KET{0}^{\! \otimes n}$, leaving the second in  $\KET{\psi_{\! s}}$.} 
\label{circuit figure}
\end{center}
\end{figure}

 The quantum circuit is shown in Figure~\ref{circuit figure}. The first step is to prepare a uniform superposition $ \KET{+}^{\otimes n} \KET{0}$ of classical states  in the first register by 
 applying Hadamard gates  ${\sf H}^{\otimes n} \! $
 to the input $ \KET{0}^{\otimes n} \! .$
 Here $\KET{+} = {\sf H} \KET{0}  =  2^{-\frac{1}{2}}(\KET{0}  \! + \! \KET{1} ).$  Applying $U$ then leads to
\begin{eqnarray}
\frac{1}{\sqrt{2^n}}
\sum_{x \in \{0,1\}^n} \KET{x} \KET{f(x)}.
\label{apply u}
 \end{eqnarray}
 To encode $s$ in the second register, note that 
 $ \sum_{x } \KET{f(x)} = (2^n \! - \! s) \KET{0} + s \KET{1} $, producing the desired result. 
 However to obtain this from (\ref{apply u})
we would have to somehow remove the classical state $\KET{x} $ in the first register. This is accomplished by a second set of Hadamards followed by postselection to $\KET{0}^{\otimes n} \! ,$ meaning the circuit of Figure~\ref{circuit figure} is repeated until the first register is measured to be $\KET{0}^{\otimes n}$ (this occurs with probability $\ge \frac{1}{2}$). But this postselection is equivalent to multiplying 
(\ref{apply u}) by $\BRA{+}^{\otimes n} \otimes I$
(and normalizing). Because 
$\BRA{+}^{\otimes n}  \KET{x} = 2^{- \frac{n}{2}}$
is a constant, independent of $x$, the first register is effectively removed from the summation, as desired.
The final result after successful postselection is therefore
\begin{equation}
\KET{\psi_{\! s}} := 
\frac{ (2^n-s) \KET{0} + s \KET{1} }{ \sqrt{ (2^n - s)^2 + s^2} }  = \cos( \textstyle{\frac{\theta_{\! s}}{2}} ) \,  \KET{0} 
+ \sin( \textstyle{\frac{\theta_{\! s}}{2}} ) \,  \KET{1},  
\ \ s \in \{ 0, 1, 2 , \cdots  , 2^n \},
\label{necklace states}
\end{equation}
where 
\begin{equation}
\theta_{\! s} =  2 \arctan ( \textstyle{ \frac{s}{2^n - s} } ), \ \ \theta_{2^n}  = \pi.
\end{equation}
The states (\ref{necklace states}), which are the potential outputs of the circuit of Figure~\ref{circuit figure}, are the inputs to the quantum state discrimination gates discussed below.
The Bloch coordinates of these initial states are
\begin{equation}
{\vec r}_s = \BRA{\psi_{\! s} } {\vec \sigma}  \KET{\psi_{\! s}} 
= ( \sin\theta_{\! s} , 0 , \cos\theta_{\! s} ) .
\end{equation}

Up to this point we have assumed the standard model of (linear) quantum computation. It should be emphasized that the circuit of Figure~\ref{circuit figure} does not allow one to efficiently measure $s$, or even determine whether $s$ is zero or nonzero. To understand why, consider a simplified 
Boolean satisfiability problem where we are promised that 
$s \! = \! 0 \ {\rm or} \ 1 $: This is the setting for UNIQUE SAT discussed below.  Measuring $s$ in this case requires discriminating between the potential states
\begin{equation}
\KET{\psi_0} = \KET{0} \ \ {\rm and} \ \ 
\KET{\psi_1} =  \cos( \textstyle{\frac{\theta_{1}}{2}} ) \,  \KET{0} 
+ \sin( \textstyle{\frac{\theta_{1}}{2}} ) \,  \KET{1}, 
\ \ \theta_1 =  2 \arctan ( \textstyle{ \frac{1}{2^n - 1} }) \approx 2^{1-n},
\label{unique sat inputs}
\end{equation}
where the approximation assumes $n \gg 1$.
However these states are exponentially close:
Their overlap  $\langle 0 \KET{\psi_1} 
= \cos( \theta_1 / 2) \approx 1 - 2^{-(2n+1)}$
is exponentially close to unity.
Discrimination protocols based on linear gates require exponential resources in this case \cite{08101970,12042313,BaeJPA15,210812299}. 
In the remainder of this paper we consider an
extended model of quantum computation
that uses nonlinear Hamiltonians and gates 
to efficiently measure $s$. 

\section{Torsion model}
\label{torsion model section}

\begin{figure}
\begin{center}
\includegraphics[width=14.0cm]{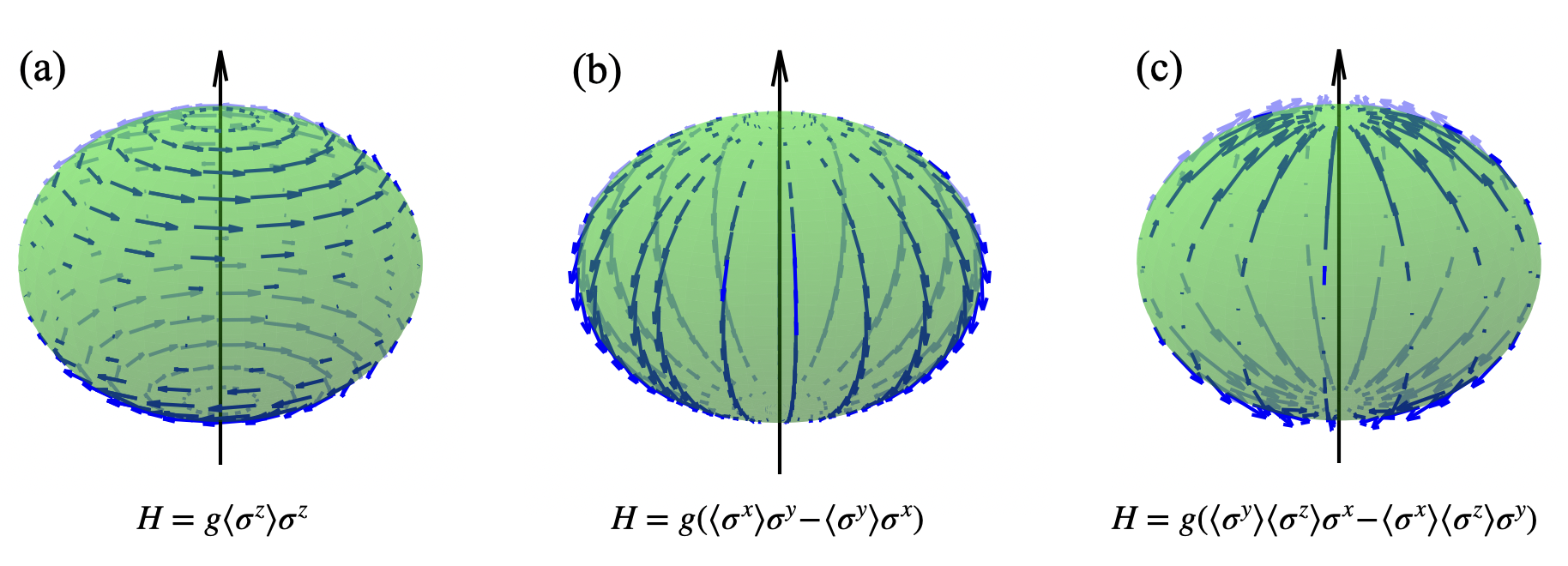} 
\caption{Nonlinear Hamiltonians and their resulting velocity fields. (a) Torsion model. (b) Morse-Smale model. (c) Supercritical pitchfork model.} 
\label{flow figure}
\end{center}
\end{figure}

A simple model for single-qubit nonlinearity is the $z$-axis torsion model 
\cite{MielnikJMP80,150706334,241022032}
\begin{eqnarray}
{\vec v} = 2gz {\vec e}_z \times {\vec r} , \ \ 
{\vec u} = 2gz {\vec e}_z , \ \ 
{\vec u} \cdot  {\vec r} \neq 0 , \ \ 
H = g \langle \sigma^z \rangle \, \sigma^z.
\label{pure torsion}
\end{eqnarray}
This model generates $z$ rotation with a frequency $2 g \langle \sigma^z \rangle $ proportional to the $z$ coordinate of the Bloch vector, as illustrated in Figure \ref{flow figure}a. The states in the upper hemisphere ($z>0$)
rotate in the direction opposite  to those in the lower hemisphere ($z<0$). Pairs of such states can move apart from each other and violate trace distance monotonicity (\ref{trace monotonicity}). Torsion (also called one-axis twist) about $x$ and $y$ axes are  similarly generated by  $ \langle \sigma^x \rangle  \sigma^x $ and  $ \langle \sigma^y \rangle  \sigma^y $. In Refs.~\cite{240416288,241022032} it was argued that $z$ axis torsion can be simulated with ultracold atoms.

Torsion on its own is not useful for state discrimination because, like any $z$ rotation,  it only causes a change of phase that does not affect measurement probabilities in the $\{ \KET{0} , \KET{1}\}$ basis. Therefore we add an $x$ rotation term to (\ref{pure torsion}):
\begin{eqnarray}
{\vec v} = 2gz {\vec e}_z \times {\vec r}
+ 2 B {\vec e}_x \times  {\vec r}  , \ \ 
{\vec u} = 2gz {\vec e}_z
+ 2 B {\vec e}_x   , \ \ 
{\vec u} \cdot {\vec r} \neq 0 , \ \ 
H = B \sigma^x + g \langle \sigma^z \rangle \, \sigma^z.
\label{torsion model}
\end{eqnarray}
In this model $\nabla \times {\vec u} = 0 $, leading to the conserved energy
\begin{eqnarray}
E = Bx + \frac{g}{2} z^2 , \ \ \nabla E = (B, 0, gz) = \frac{ {\vec u}}{2}.
\end{eqnarray}
Note that the conserved energy is not   equal to $ \langle H \rangle = Bx + g z^2$ [recall (\ref{energy conservation linear case})], but is instead given by $ E = \langle H \rangle - \frac{g}{2} z^2$, a common feature in mean field theories \cite{PethickSmith2008,250306423}.

\subsection{Quantum state discrimination and UNIQUE SAT}

The torsion model (\ref{torsion model}) has been  investigated  previously by Childs and Young \cite{150706334}, and by Geller \cite{240416288,241022032}.
Childs and Young \cite{150706334} constructed an efficient  quantum state discrimination gate by using a time-dependent $B$ to maximally increase the trace distance between two potential inputs 
(for example $ \KET{\psi_0} , \KET{\psi_1} $)
until the states orthogonalize. In Ref.~\cite{241022032}, the setting $B \! = \! \frac{g}{2}$ was used to construct an efficient quantum state discrimination gate where the qubit  trajectories follow segments of Viviani's curve at the intersection of the Bloch sphere and a cylinder tangent to the sphere.  Here we consider an even simpler variation adapted to the efficient solution of UNIQUE SAT \cite{BlassGurevichIC1982}. In this {\it promise} problem, we are given a Boolean formula $f(x)$ and a promise that the number of solutions $s$ is $0 \! \le \!  s \!  \le \! 1$, and the objective is to determine if $ s\! = \!1$ (if \textsc{False},  we know $s=0$). This problem is distinct from the UNIQUE SAT {\it decision} problem where $ 0 \le s \le 2^n$ 
and the objective again is to determine if $ s\! = \!1$. The Valiant–Vazirani theorem \cite{ValiantVaziraniTCS1986} established that 
SAT reduces to UNIQUE SAT under a randomized polynomial time reduction. Therefore
the UNIQUE SAT promise problem is {\sf NP}-hard under randomized polynomial-time reduction.
Although not proven, it is believed that 
UNIQUE SAT is not contained in the complexity class {\sf BPP} (bounded error probabilistic polynomial time), making it beyond the reach of efficient classical computation.
The state discrimination gate derived here is of interest because of its simplicity and the ease of extending its energy--based construction to other applications.
The gates of 
Refs.~\cite{150706334,241022032} would also enable
efficient UNIQUE SAT solution.

The first step of the gate is to rotate the inputs
$\{ \KET{\psi_0} , \KET{\psi_1}\} $ from (\ref{unique sat inputs}) to $\{ \KET{a} , \KET{b} \} $ by applying a $y$ rotation 
$R_y(\gamma)  =  e^{- i\frac{\gamma}{2}  \sigma^y} \! , $ where $\gamma = \frac{\pi}{2}  - \frac{\theta_1}{2}$.  The rotated states $\{ \KET{a} , \KET{b} \} $ have Bloch vectors
\begin{eqnarray}
{\vec r}_a = \textstyle{ \left(  \sin(\frac{\pi}{2}-\frac{\theta_1}{2}) , 0 , \cos(\frac{\pi}{2}-\frac{\theta_1}{2}) \right) 
=   \left( \cos \frac{\theta_1}{2}  , 0 , \sin \frac{\theta_1}{2} \right) }
\end{eqnarray}
and
\begin{eqnarray}
{\vec r}_b = \textstyle{ \left( \sin(\frac{\pi}{2}+\frac{\theta_1}{2}) , 0 , \cos(\frac{\pi}{2}+\frac{\theta_1}{2}) \right)
=  \left( \cos \frac{\theta_1}{2} , 0 , - \sin \frac{\theta_1}{2} \right) } ,
\end{eqnarray}
straddling the equator in the $y=0$ plane.
When $n$ is large, the angle $\theta_1 =  2 \arctan ( \textstyle{ \frac{1}{2^n - 1} }) 
\approx 2^{1-n} $ between these Bloch vectors is exponentially small. We seek trajectories mapping ${\vec r}_a$ to
the north pole ${\vec r} = (0,0,1)$ and  ${\vec r}_b$ to  the south pole ${\vec r} = (0,0,-1)$.
The rotated states both have initial energy
$ E_{\rm i} = B \cos \! \left(\frac{\theta_1}{2}\right)  + \frac{g}{2} 
\sin^2 \! \left(\frac{\theta_1}{2}\right)$.
The final states at the poles  have energy
$ E_{\rm pole} =  \frac{g}{2} $.
To equate these, choose 
\begin{eqnarray}
B =  \frac{g}{2}  \cos(\frac{\theta_1}{2}) .
\end{eqnarray}

In contrast to \cite{241022032}, $B \neq \frac{g}{2}$ and the states here do not follow Viviani curves.
To find the trajectories, use energy conservation to write $x$ and $y$ in terms of $z$.
Using
\begin{eqnarray}
B x + \frac{g}{2} z^2 = E_{\rm pole}  
\end{eqnarray}
leads to
\begin{eqnarray}
x =  \frac{1-z^2}{ \cos( \frac{\theta_1}{2})  } 
\ \ {\rm and} \ \ 
y^2 =  1 - x^2 - z^2 =  
\frac{ (1-z^2)  [z^2 - \sin^2( \frac{\theta_1}{2})  ] }{ \cos^2( \frac{\theta_1}{2}) } .
\label{x and y elimination}
\end{eqnarray}
From the velocity field  in (\ref{torsion model}) we 
have
\begin{eqnarray}
\frac{dz}{dt} = 2 B y =  g y \cos( \frac{\theta_1}{2}) 
= \pm g \sqrt{ \bigg( \! 1-z^2 \bigg) \!  \left(z^2 -   \sin^2 \frac{\theta_1}{2}   \right)} .
\end{eqnarray}
The $\pm$ gives the two possible signs of $y$ 
after taking the square root in (\ref{x and y elimination}). The positive $y$ case applies to the trajectory in the upper hemisphere with $\frac{dz}{dt} > 0$, mapping ${\vec r}_a$  to the north pole. 
The $\frac{dz}{dt} < 0$ case maps ${\vec r}_b$  to the south pole. The evolution times are the same.

\subsection{Time complexity}

The gate time for ${\vec r}_a$  to reach the pole at $z=1$ is
\begin{eqnarray}
t_{\rm g} = \frac{1}{g} 
 \int_{\sin \! \frac{\theta_1}{2} }^{1}
\frac{dz}{  \sqrt{ \big( \! 1-z^2 \big) \!  \left(z^2 -   \sin^2 \frac{\theta_1}{2}   \right)} }  .
\end{eqnarray}
The substitution
$ z^2 = \sin^2( \frac{\theta_1}{2}) + 
\cos^2 (\frac{\theta_1}{2})  \sin^2 \! w $
leads to
\begin{eqnarray}
t_{\rm g}  = 
\frac{1}{g \sin \frac{\theta_1}{2} }  \int_{0}^{\frac{\pi}{2}}
\frac{dw}{\sqrt{1 + \cot^2( \frac{\theta_1}{2}) \sin^2 (w)}} 
= \frac{ {\sf K}( i \cot \frac{\theta_1}{2} ) }{g \sin \frac{\theta_1}{2} } ,
\label{gate time elliptical integral}
\end{eqnarray}
where
\begin{eqnarray}
{\sf K}(k)  := \int_{0}^\frac{\pi}{2}
\frac{dw}{\sqrt{ 1-k^2  \sin^2w }} 
\label{k definition}
\end{eqnarray}
is the complete elliptic integral of the first kind.
We evaluate (\ref{k definition}) for imaginary argument
$ k =  i \cot \frac{\theta_1}{2}$ by
using a standard  identity 
$ {\sf K}( i \frac{q}{ \sqrt{ 1-q^2}}   ) 
= \sqrt{1-q^2} \,  {\sf K}(q)$ \cite{Gradshteyn2000},
valid for any $q \in (0,1)$. 
Let $ \lambda  :=  \frac{q}{ \sqrt{ 1-q^2}} $. The identity leads to
\begin{eqnarray}
{\sf K}(i\lambda) =  \frac{  {\sf K}( \frac{\lambda}{\sqrt{1+\lambda^2}} )  }{ \sqrt{1+\lambda^2} } , \ \ \lambda > 0.
\end{eqnarray}
Then (\ref{gate time elliptical integral}) becomes
\begin{eqnarray}
t_{\rm g}  = \frac{1}{g} \,
 {\sf K}(  \cos {\textstyle { \frac{\theta_1}{2} }} ) 
\approx 
\frac{1}{g} \log \! \left( \frac{8}{ \theta_1 } \right) \! ,
 \label{gate time simplified}
\end{eqnarray}
where the approximation assumes large $n$
(small $\theta_1$). The gate time $ t_{\rm g} = O(n) $ is polynomial in $n$ as required.

\section{Morse–Smale flow on $S^2$}
\label{morse–smale section}

Next we design a nonlinear Hamiltonian and gate to solve 3SAT \cite{AroraBarak2009}.
The setting is the same as above: $f(x)$ is a Boolean function on $n$ bits in conjunctive normal form with $s$ satisfying assignments. However  
now
$ 0 \le s \le 2^n$ 
and the objective is 
to find whether $s$ is zero or nonzero.
To achieve this we create an unstable fixed point (source) at $ {\vec r} = (0,0,1) = {\vec e}_z $  and a stable  fixed point (sink) at
$ {\vec r} = (0,0,-1) =  -{\vec e}_z $. 
Then, every input  other than  $\KET{0}$, the unstable fixed point, will flow to $\KET{1}$ at long times. To find a suitable velocity field, note that   $ {\vec v} = -  {\vec e}_z $ gives a downward flow  $( dz/dt < 0)$, but is not tangential. To make it tangential we can use
 \begin{equation}
{\vec v} =  2g( -  {\vec e}_z  + z {\vec r}  )
=  2g( xz, yz, z^2-1), \ \ g > 0,
\label{morse–smale flow}
\end{equation}
which satisfies both 
${\vec v} \cdot {\vec r} = 0$ and
\begin{equation}
\frac{dz}{dt} = 2g(z^2 - 1)  \le 0 .
\label{dzdt morse–smale flow}
\end{equation}
The flow is illustrated in Figure \ref{flow figure}b. 
Here we have included a scale factor $2g$.
Solving  (\ref{dzdt morse–smale flow}),  the height evolves as
\begin{equation}
z(t) = (1+c e^{4gt}) (1- c e^{4gt})^{-1} \! ,  \ \ 
{\rm where} \ \ 
c = (z(0) \!  - \!  1) (z(0)\!  + \! 1)^{-1} \! .
\label{morse–smale height}
\end{equation}

The velocity field (\ref{morse–smale flow}) has two fixed points 
$  {\vec r} = \pm {\vec e}_z,$
as desired. It is a standard example of 
Morse–Smale flow \cite{HirschSmale2013} on a compact manifold, with a source at the north pole and a sink at the south pole.  Due to the azimuthal  symmetry of the velocity field, the dynamics foliates into 1d curves with fixed longitude (azimuthal angle $\phi$).
All trajectories connect the poles and  there is no recurrence or chaos. In addition, Morse–Smale flow is structurally stable under perturbation, meaning that small changes in the model leave the qualitative dynamics unchanged \cite{HirschSmale2013}. 
To find the Hamiltonian, note that
\begin{equation}
{\vec u}= {\vec r}  \times {\vec v} =  2g \, {\vec e}_z \times {\vec r} =   2g \, (-y, x, 0), \ \ 
{\vec u} \cdot {\vec r} = 0.
\label{morse–smale u}
\end{equation}
The nonlinear Hamiltonian for the Morse–Smale model is
\begin{equation}
H = g \bigg( \! \langle \sigma^x \rangle \sigma^y -  \langle \sigma^y \rangle \sigma^x \bigg).
\label{hamiltonian morse–smale flow}
\end{equation}
In this model  $ {\rm curl}_S \, {\vec u} \neq  0$
and energy is {\it not} conserved.
Energy non-conservation is discussed further in Section \ref{conclusion section}.

\subsection{Quantum state discrimination and 3SAT}

Having obtained the desired nonlinear Hamiltonian, we use the velocity field (\ref{morse–smale flow}) to design a quantum state discrimination gate generated by this Hamiltonian.
The gate works by applying the Hamiltonian
(\ref{hamiltonian morse–smale flow})
 for a time $t_{\rm g}$. All initial states 
$\KET{\psi_s}$ other than 
$\KET{\psi_0} = \KET{0} $ will flow to $\KET{1}$. Subsequent measurement of the qubit indicates whether  $s \! = \! 0$ (the Boolean formula is not satisfiable) or $s \! > \! 0$ (the formula is satisfiable).  The 
 heights satisfy
\begin{equation}
 z_0 >  z_1 > z_2 > \cdots > z_{2^n} , 
\end{equation}
throughout the evolution $t \ge 0.$
Thus, the gate time is determined by the trajectory of the $s \! = \! 1$  state
$\KET{\psi_1}$, which has initial angle
$ \theta_1 =  2 \arctan ( \textstyle{ \frac{1}{2^n - 1} }) \approx 2^{1-n} $
and
initial height  $ z_{\rm i} = \cos(\theta_1) \approx 
 1 - 2^{1-2n} $, where the approximations assume $n \gg 1$.
 
 \subsection{Time complexity}
 
To find the gate time, use
(\ref{dzdt morse–smale flow})  to obtain 
\begin{equation}
 t_{\rm g} = 
 \frac{1}{2g} \int_{z_{\rm i}}^{z_{\rm f}}  \frac{dz}{ z^2 - 1}  =
  \frac{1}{4g} \bigg[ \log( \bigg|  \frac{ z-1}{z+1}  \bigg| ) \bigg]_{z_{\rm i}}^{z_{\rm f}}  \! \! .
\label{gate time general formula}
\end{equation}
Ideally we would like to set $z_{\rm f} = -1$, which would map all $s \! > \! 0$ states to $\KET{1}$.
However we can see from (\ref{gate time general formula}) that this would take infinite time. 
However, these states do become exponentially close to  $\KET{1}$ in polynomial time. Therefore we introduce a small error parameter $\epsilon $ that bounds the final $z$ coordinates according to
\begin{equation}
z_{\rm f} = -1 + \epsilon, \ \ 0 < \epsilon \ll  1.
\label{almost pole}
\end{equation}
Then we obtain
\begin{equation}
t_{\rm g} = \frac{1}{4g}  \log( \frac{ (2- \epsilon) 
(1 - 2^{-2n} )  } { 2^{-2n}  \, \epsilon     } ) \approx
\frac{1}{4g}  \log( \!  \frac{2^{2n+1} }{\epsilon}  ),
\end{equation}
where the approximation applies to small $\epsilon$ and large $n$.
The gate is efficient even for exponentially small 
$\epsilon$.

\section{Bistable flow on $S^2$ with two attracting fixed points}
\label{pitchfork section}

Finally, we design a nonlinear Hamiltonian  to measure the precise value of $s$ using
Aaronson's binary search \cite{0502072}.
The algorithm requires $n+1$ applications of a quantum state discrimination gate that generates bistable flow with attractive fixed points at the poles $\KET{0}$ and $\KET{1}$. We seek  a velocity field that divides the Bloch ball into two basins of attraction, the upper and lower hemispheres, with a separatrix at the equatorial plane.
This is accomplished with the velocity field
 \begin{equation}
{\vec v}=  2gz(   {\vec e}_z  -  z {\vec r}  )
=  2g ( -x z^2 , -y z^2, z - z^3), \ \ g > 0,
\label{pitchfork flow}
\end{equation}
with fixed points at
\begin{equation}
 {\vec r} = (0,0, \pm 1) \ {\rm (attractive)}
 \ \ {\rm and \ the \ equator}  \ z=0 \ {\rm (repulsive)}  .
\end{equation}
The flow is illustrated in Figure \ref{flow figure}c. 
This field is obtained by multiplying (\ref{morse–smale flow}) by $-z$, which changes the stability of the fixed point at $\KET{0}$ and creates a new line of unstable 
fixed points  on the equator.
The $z$ coordinate now evolves according to
\begin{equation}
\frac{d z}{dt} = 2g z(1-z^2),
\label{pitchfork equation}
\end{equation}
having the dynamics  of a supercritical pitchfork bifurcation \cite{StrogatzNonlinearDynamics2015} in the unstable phase. In particular,  (\ref{pitchfork equation}) describes a bistable 1d gradient flow with two stable equilibria at $z=\pm1$ separated by an unstable equilibrium at $z=0$. 
Solving (\ref{pitchfork equation}) 
in the upper hemisphere, 
the height evolves as
\begin{equation}
z(t) = (1+ c e^{-4gt})^{-\frac{1}{2}} \! ,  \ \ 
{\rm where} \ \ 
c = z(0)^{-2} - 1 .
\label{pitchfork height}
\end{equation}
(Note, however, that the Bloch vector does not follow the $z$ axis to a pole but remains on the  sphere ${|\vec r}| = 1$.)  As with the Morse-Smale flow, the dynamics foliates into 1d curves with fixed longitude. The velocity field (\ref{pitchfork flow}) leads to a ${\vec u}$ field (\ref{u field}) given by
\begin{equation}
{\vec u}  =
2 g z \, {\vec r} \times {\vec e}_z
=  2g \, (yz, -xz, 0), \ \ 
{\vec u} \cdot {\vec r}  = 0,
\label{pitchfork u}
\end{equation}
with  
$ {\rm curl}_S \, {\vec u} \neq 0 $	.
The  nonlinear Hamiltonian for the supercritical pitchfork model is
\begin{equation}
H = g  \bigg( \! \langle \sigma^y \rangle
 \langle \sigma^z \rangle  \sigma^x
  -  \langle \sigma^x \rangle \langle \sigma^z \rangle \sigma^y  \bigg).
\label{hamiltonian pitchfork flow}
\end{equation}
This is another example of energy non-conservation due to the presence of sources and sinks in the velocity field. 

\subsection{Quantum state discrimination and \#SAT}

To find the time required for each application of the quantum state discrimination gate, use (\ref{pitchfork equation})
to obtain
\begin{equation}
t_{\rm g} =  \frac{1}{2g}  \int_{z_{\rm i}}^{z_{\rm f}} \frac{dz}{z-z^3}=  \frac{1}{2g} 
\bigg[ \log( \bigg| \frac{z}{ \sqrt{1-z^2}} \bigg| )
\bigg]_{z_{\rm i}}^{z_{\rm f}}  .
\end{equation}
In this case we let $z_{\rm i} \rightarrow 0 $ and
$z_{\rm f} \rightarrow 1 - \epsilon$, where
$  \epsilon > 0$ is a small error in the final state.
Then 
\begin{equation}
t_{\rm g} =  \frac{1}{2g} 
\log(  \frac{1}{ \sqrt{2 \epsilon \, } z_{\rm i} } ) ,
\label{pitchfork tg}
\end{equation}
which is efficient even for exponentially small 
$ \epsilon $ and $ z_{\rm i} $.
The gate time in the lower hemisphere is the same.

\subsection{Binary search}

To count the number of solutions $ s \in \{ 0, 1, \cdots , 2^n \}$, we express $s$ as an $n+1$-bit binary number
\begin{equation}
s = (s_{n} \cdots s_1 s_0) = \sum_{j=0}^{n} s_j \, 2^j,  \ \  s_j \in \{0,1\},
\end{equation}
and measure each bit $s_j$ using binary search \cite{0502072},
starting with the most significant bit $s_{n}$.
More precisely, the measurement of 
$s_{n-1} \cdots s_1 s_0$ uses binary search 
but the measurement  of $s_n$ follows a modified protocol. To measure $s_n$, 
apply a negative $y$ rotation by an angle 
$\gamma = ( \frac{ \theta_{2^n - 1} + \theta_{2^n} 
}{2} ) -  \frac{\pi}{2}$ 
to move a plane of separation on the Bloch sphere,  at polar angle
$  ( \frac{ \theta_{2^n - 1} + \theta_{2^n} 
}{2} ), $ to the 
equatorial plane (separatrix).
This leaves states
 $ s \in \{ 0, 1, \cdots , 2^n - 1\}$
in the upper hemisphere and 
$ s \in \{ 2^n \}$ in the lower hemisphere. 
Apply the Hamiltonian 
(\ref{hamiltonian pitchfork flow}) for a time $t_{\rm g}$, followed by measurement of the nonlinear qubit in the $\{ \KET{0} , \KET{1} \} $ basis. The measurement outcome is 
$  s_n \in \{0,1\} $. If $s_n \! = \! 1$, return
$ s = (1 0 \cdots 0) = 2^n$ and halt
(no further operations are needed).

If $s_n \! = \! 0$,  initialize an interval $ [s_{\min} , s_{\max} ] := \{ s \in {\mathbb Z} :  s_{\min}  \le s  \le s_{\max} \} $ 
containing possible $s$ values to
$ [ 0 , 2^{n} -1 ] $. Let $N$ be the size (number of integers) of $[s_{\min} , s_{\max} ]$. $N$ is initially 
equal to $2^n$, and it  remains a power of 2 under bisection, until $N \! = \! 1$ is reached.
Let $j = n-1$ and repeat the following procedure $n$ times:
 \begin{enumerate}
 
\item Divide the current interval into equal halves 
$ [s_{\min} , s_{\max} ]  \mapsto   [s_{\min} , s_{k} ] \cup [s_{k+1} , s_{\max} ] $, 
 where $ k  = \lfloor \frac{ s_{\min} + s_{\max} }{2}  \rfloor $.  Here $\lfloor x \rfloor$  is the greatest integer less than or equal to $x $.
Introduce a plane separating these two sets
on the Bloch sphere. The plane passes through the origin and has polar angle $\frac{  \theta_{k} + \theta_{k+1}}{2} $, at the midpoint between states with $s=k$ and $s=k+1$.

 \item Apply a positive  $y$ rotation  by an angle
 $ \gamma = \frac{\pi}{2} - (\frac{\theta_k + \theta_{k+1}}{2})$, 
 rotating the plane of separation to the equatorial plane. This places  the inputs (\ref{necklace states}) with
 $ s \in [s_{\min} , s_{k} ] $ in the upper hemisphere, and those with $ s \in  [s_{k+1} , s_{\max} ] $ in the lower hemisphere.
 
\item Apply the Hamiltonian 
(\ref{hamiltonian pitchfork flow}) for a time $t_{\rm g}$ and measure 
 in the $\{ \KET{0} , \KET{1} \} $ basis,
recording the result as $ s_j \in  \{ 0,1\} $.
If $s_j=0$, decrease the upper bound: $ s_{\max}  \leftarrow s_{k}$.  If $s_j=1$, increase the lower bound: $ s_{\min}  \leftarrow s_{k+1}$.

\item Return $s_{j}$, halve $ N \leftarrow  \frac{N}{2}$,  and decrement $j \leftarrow j -1$.

 \end{enumerate}
 
 \subsection{Time complexity}
 
 After repeating the procedure $n$ times, the number of solutions is given in binary form as $ s = (s_{n} \cdots s_1 s_0)$.
 The total algorithm duration is $n+1$ times the gate time (\ref{pitchfork tg}), 
 with $z_{\rm i} = O(2^{-n})$. 
Therefore, the nonlinear quantum algorithm solves \#{SAT} with time complexity
 \begin{equation}
t = \frac{1}{g} \,
O \left( n \log(  2^n  \epsilon^{-\frac{1}{2}}   ) 
\right).
\end{equation}
This would provide, 
in the noise-free limit, an efficient solution of 
\#{SAT}, which is a \#{\sf P}-complete problem.

\section{Conclusions}
\label{conclusion section}

In this work, we  study hypothetical forms of 
single-qubit nonlinearity and their resulting computational power in the noise-free limit.
In addition to the torsion model, which has been investigated previously \cite{MielnikJMP80,150706334,241022032}, we study two models with engineered phase portraits. 
Building on Refs.~\cite{150706334} and \cite{241022032} we argue that a Hamiltonian with  $ \langle \sigma^z \rangle \sigma^z$ torsion nonlinearity can solve UNIQUE SAT in polynomial time. Valiant and Vazirani proved that the complexity class of a UNIQUE SAT solver with access to randomness is {\sf NP}-hard \cite{ValiantVaziraniTCS1986}.
Second, we propose to use a Hamiltonian with 
$ \langle \sigma^x \rangle \sigma^y 
\! - \! \langle \sigma^y \rangle \sigma^x$ nonlinearity, which we call the Morse–Smale model, to solve 3SAT in polynomial time.
As the first {\sf NP}-complete problem, 3SAT plays an especially prominent role in the theory of computing \cite{AroraBarak2009}.
Third, we argue that
$ \langle \sigma^y \rangle
\langle \sigma^z \rangle  \sigma^x
\! - \! \langle \sigma^x \rangle
\langle \sigma^z \rangle  \sigma^y $ 
nonlinearity, the supercritical pitchfork model, can be used to solve \#SAT, which is \#{\sf P}-complete.

It's intriguing that a single ancilla qubit evolving nonlinearly confers so much power to a scalable fault-tolerant  quantum computer \cite{PhysRevLett.81.3992}. But it is also important to emphasize the limitations of these results.
One limitation is that any discussion of complexity necessarily assumes a scalable, fault-tolerant implementation, which is not addressed here. 
And how can we access the required nonlinearity?

The nonlinear Hamiltonians are of mean field type, depending on the current state 
$ \KET{\psi} $ through  $\langle \sigma^\mu \rangle = \BRA{\psi} \sigma^\mu \KET{\psi} $.
This suggests that they might be simulated by quantum many-body systems in some mean field regime. Previous work suggests that
 $ \langle \sigma^z \rangle \sigma^z $ nonlinearity 
 ($z$ axis torsion) can be simulated with ultracold atoms \cite{240416288,241022032}.
The Morse–Smale and supercritical pitchfork models with 
$ \langle \sigma^x \rangle \sigma^y 
\! - \! \langle \sigma^y \rangle \sigma^x$ 
and
$ \langle \sigma^y \rangle
\langle \sigma^z \rangle  \sigma^x
\! - \! \langle \sigma^x \rangle
\langle \sigma^z \rangle  \sigma^y $ 
nonlinearity appear to be somewhat exotic and don't conserve energy. This suggests that the models are emergent in nature.  In particular, they might arise as  effective Hamiltonians for microscopically time-dependent systems, such as Floquet effective Hamiltonians.
These issues deserve further investigation.

\acknowledgements

This work was partly supported by the NSF under grant no.~DGE-2152159.

\appendix


\bibliographystyle{unsrtnat}
\ifUSEBBL
\bibliography{MS2.bbl}
\else
\bibliography{/Users/mgeller/Dropbox/bibliographies/CM,/Users/mgeller/Dropbox/bibliographies/MATH,/Users/mgeller/Dropbox/bibliographies/QFT,/Users/mgeller/Dropbox/bibliographies/QI,/Users/mgeller/Dropbox/bibliographies/group,/Users/mgeller/Dropbox/bibliographies/books}
\fi

\end{document}